\documentstyle[12pt,a4wide]{article}
\begin{document}
\input epsf
\newcommand{\be}{\begin{equation}}
\newcommand{\ee}{\end{equation}}
\newcommand{\pr}{\partial}
\newcommand{\ie}{{\it ie }}
%%%%%%%If you do not have the msbm fonts, delete the following 10 lines
\font\mybb=msbm10 at 11pt
\font\mybbb=msbm10 at 17pt
\def\bb#1{\hbox{\mybb#1}}
\def\bbb#1{\hbox{\mybbb#1}}
\def\Z {\bb{Z}}
\def\R {\bb{R}}
\def\E {\bb{E}}
\def\T {\bb{T}}
\def\C {\bb{C}}
%%%%%%%%%%%%%%%%%
\renewcommand{\theequation}{\arabic{section}.\arabic{equation}}
\newcommand{\news}{\setcounter{equation}{0}}
\newcommand{\bx}{{\bf x}}
\newcommand{\cmon}{{\cal C}^{\mbox{mon}}}
\newcommand{\smon}{{\cal C}^{\mbox{sky}}}
\newcommand{\bpi}{\mbox{\boldmath $\pi$}}     
\newcommand{\pauli}{\mbox{\boldmath $\tau$}}   
\newcommand{\bpsi}{\mbox{\boldmath $\psi$}}
     
\title{\vskip -70pt
\begin{flushright}
%{\normalsize \sl Draft } \\
\end{flushright}\vskip 50pt
{\bf \large SPHALERONS IN THE SKYRME MODEL \bf }\\[30pt]
\author{Steffen Krusch and Paul Sutcliffe
\\[10pt]
\\{\normalsize   {\sl Institute of Mathematics,}}
\\{\normalsize {\sl University of Kent,}}\\
{\normalsize {\sl Canterbury, CT2 7NZ, U.K.}}\\[10pt]
{\normalsize{\sl Email : S.Krusch@kent.ac.uk}}\\
{\normalsize{\sl Email : P.M.Sutcliffe@kent.ac.uk}}\\[20pt]
\normalsize{PACS: 11.10.Lm}\\}}
\date{June 2004}
\maketitle

\begin{abstract}
\noindent 
Numerical methods are used to compute sphaleron solutions
of the Skyrme model. These solutions have topological 
charge zero and are axially symmetric,
consisting of an axial charge $n$ Skyrmion and an axial
 charge $-n$ antiSkyrmion 
(with $n>1$), balanced
in unstable equilibrium. The energy is slightly less than twice the
energy of the axially symmetric charge $n$ Skyrmion. 
A similar configuration with $n=1$ does not produce
a sphaleron solution, and this difference is explained by considering
the interaction of asymptotic pion dipole fields. 
For sphaleron solutions with $n>4$ the
positions of the Skyrmion and antiSkyrmion merge to form a circle, 
rather than isolated points, and there are some features in
common with Hopf solitons of the Skyrme-Faddeev model.

\end{abstract}
\newpage

\section{Introduction}\news\label{sec-intro}
Sphalerons are unstable classical solutions of field theories whose
existence is due to non-trivial topological properties of the space of field
configurations. Taubes was the pioneer of a topological approach to
 finding saddle point solutions and used this method to prove the
existence of a monopole-antimonopole solution of the Yang-Mills-Higgs
equations \cite{Tau}. Subsequently, Manton used a similar approach to 
suggest the existence of an unstable solution in the Weinberg-Salam 
theory \cite{Ma9},
which was then studied numerically in ref.\cite{KlMa} where the name 
sphaleron was coined. These, and other, successes prompted the search
for a sphaleron in the Skyrme model, corresponding to a
Skyrmion-antiSkyrmion solution in analogy with the monopole-antimonopole 
solution of the Yang-Mills-Higgs theory. However, despite the fact
that the topological aspects are similar in the two theories, the 
current evidence \cite{BGS,ILP,GP} suggests that such a sphaleron 
is unlikely to exist, although it remains an open problem.

In this paper we observe that the asymptotic pion dipole interactions
between Skyrmions and antiSkyrmions suggests that a more promising
candidate for a saddle point solution consists of an axially
symmetric charge 
$n$ Skyrmion balanced
in unstable equilibrium with a charge $-n$ antiSkyrmion, where $n>1$ 
rather than $n=1.$ Note that for $n>2$ the constituents are themselves
already saddle point solutions, since the minimal energy charge $n$
Skyrmion is not axially symmetric for $n>2$ \cite{BS3}.

Using numerical
simulations of the Skyrme model we investigate this possibility and
indeed are able to compute static saddle point solutions. 
Examples with $2\le n\le 6$ are presented in detail. 
They are axially symmetric and the energy is slightly less than twice the
energy of the axial charge $n$ Skyrmion. 
Some comments on the topology associated with these solutions are made.

For $n=3$ and $n=4$
the solutions are qualitatively similar to the $n=2$ solution, with
the separation between the Skyrmion and antiSkyrmion being smaller
for larger values of $n.$ For $n>4$ the
positions of the Skyrmion and antiSkyrmion are no longer isolated, but 
merge to form a circle, producing solutions which have some features in
common with Hopf solitons of the Skyrme-Faddeev model \cite{FN,GH,BS5,HS}.

\section{Topology and Interaction Energies}\news\label{sec-dipole}
Let us begin by recalling the salient features behind the
existence of Taubes' monopole-antimonopole solution. 
Morse theory relates the topology of a manifold to the number and
types of critical points of a function defined on the manifold.
Taubes \cite{Tau} was able to apply an infinite dimensional version of Morse
theory, with the space of field configurations and the energy functional 
playing the role of the manifold and the function respectively.
Topology enters through the  non-triviality of certain homotopy groups,
as follows.
The space $\cmon$ of
 finite energy Yang-Mills-Higgs field configurations 
(after suitably removing the gauge freedom) is homotopic to the space
of maps from $S^2$ to $S^2,$ which may be thought of as the space
of Higgs fields at infinity. The degree of this map $N$ is the monopole
number and this labels the connected components of $\cmon$ since
\be
\pi_0(\cmon)=\pi_0(\mbox{Maps}(S^2\mapsto S^2))=\pi_2(S^2)=\Z.
\ee   
The component $\cmon_0$, with zero monopole number, 
has non-contractible
loops since
\be
\pi_1(\cmon_0)
=\pi_1(\mbox{Maps}(S^2\mapsto S^2))=\pi_3(S^2)=\Z
\label{clm}\ee
where now the maps between two-spheres must be restricted to those 
with degree zero.
 
A generator for this homotopy group is the non-contractible loop
in configuration space in which a monopole-antimonopole pair is 
created from the vacuum, the
pair are separated and the monopole rotated by $2\pi$ before the
pair are brought together again to annihilate back to the vacuum.
For any loop $\gamma$ in the same homotopy class as this one we can determine 
the maximal value $E_\gamma$ of the energy along this loop. Minimizing this
maximal value over all loops $\gamma$ yields an unstable stationary point 
of the energy functional and this is the sphaleron. It is the midpoint of
the non-contractible loop where the monopole is rotated by $\pi$
and the pair relax to the optimal separation to minimize the energy
with this relative rotation. The pair can not relax to annihilate because
the loop is non-contractible and hence the only other way that this procedure
could fail to yield a sphaleron is if the relaxation produced a 
monopole-antimonopole pair with infinite separation. Taubes was able to
rigorously rule out this possibility, and physically it corresponds to the
fact that at large separation the interaction energy of a monopole-antimonopole
pair is dominated by the Coulomb force which has a magnetic and (in the
BPS limit) a scalar contribution which are both strictly attractive.
This attraction at large separation therefore prevents the relaxation from
producing a pair with infinite separation. The monopole-antimonopole
solution is axially symmetric and has been computed numerically
\cite{Rub,KlKu}, where its energy (in the BPS limit) is found to be
$1.70$ times the energy of a single monopole.

Some aspects of the above analysis can be mirrored in the Skyrme model,
though others can not, as we now briefly describe.

The space $\smon$ of
 finite energy Skyrme fields \cite{Sk} is homotopic to the space
of maps from $S^3$ to $S^3.$  The degree of the map $B$ is the baryon
number and this labels the connected components of $\smon$ since
\be
\pi_0(\smon)=\pi_0(\mbox{Maps}(S^3\mapsto S^3))=\pi_3(S^3)=\Z.
\ee   
As for monopoles, the component $\smon_0$, with zero charge, 
has non-contractible
loops since
\be
\pi_1(\smon_0)
=\pi_1(\mbox{Maps}(S^3\mapsto S^3))=\pi_4(S^3)=\Z_2
\label{clsky}\ee
where we restrict to degree zero maps between three-spheres.
Note that this result shows that in the Skyrme model there is only
one type of non-contractible loop, compared to the infinite number
for monopoles.

The non-contractible loop is generated by creating a Skyrmion-antiSkyrmion
pair, separating them, rotating the Skyrmion by $2\pi$ and bringing
the pair back together to annihilate. So far the discussion is very
close to that for monopoles, the only difference being that rotating the
Skyrmion by $4\pi$ instead of $2\pi$ is a contractible loop in the Skyrme
model, but this is not important for the possible existence of a sphaleron.
However, recall that a sphaleron could fail to exist if the minimax
procedure results in a Skyrmion-antiSkyrmion pair with infinite separation,
and this is where the crucial difference lies. For monopoles this does not
happen because a monopole-antimonopole pair attract at large separations
for all relative phases, but this is not true for a Skyrmion-antiSkyrmion pair.

From far away a Skyrmion resembles a triplet of orthogonal pion dipoles.
Let us denote the dipole strength by $4\pi C,$ where $C$ is a positive
constant. The leading order contribution to the interaction energy
of a well separated Skyrmion-antiSkyrmion pair is given by the dipole-dipole
interaction term. Consider a Skyrmion-antiSkyrmion pair with separation $s$
and the Skyrmion rotated by a 
phase\footnote{The relative phase  and the normalization
 of the energy are defined explicitly in section \ref{sec-axial}.}
 $\alpha$ relative to the antiSkyrmion, about the line joining them. 
To leading order the energy
of the pair is \cite{BGS,ILP}
\be
E=2E_1-\frac{4C^2}{3\pi s^3}(1+\cos\alpha)
\label{int1}
\ee
where $E_1$ denotes the energy of a single Skyrmion.

Providing the relative phase is not equal to $\pi$ then the interaction
energy is negative so the Skyrmion and antiSkyrmion attract. However, if
the phase is $\pi$ then the dipole-dipole interaction vanishes and 
a higher order calculation must be performed to determine the 
nature of the interaction. The result of such a calculation \cite{ILP}
reveals that the leading contribution to the interaction energy is
of order $1/s^6$ and is positive, so that a Skyrmion and antiSkyrmion
are repulsive when the relative phase is $\pi.$ 
The result of the minimax procedure is therefore a Skyrmion and
antiSkyrmion with infinite separation, not a sphaleron. The numerical
results presented in the following sections agree with this analysis and
support the conclusion that there is no sphaleron in the Skyrme model
that may be thought of as a single Skyrmion-antiSkyrmion pair.

Let us turn our attention, for the moment, 
to the minimal energy charge 2 Skyrmion.
This is axially symmetric and its asymptotic fields resemble a single
pion dipole aligned with the symmetry axis and with a dipole strength of
approximately $8\pi C.$ This is because the charge 2 Skyrmion is formed
by bringing together two single Skyrmions where one is rotated
by $\pi$ around an axis (which will become the axis of symmetry)
orthogonal to the line joining the two Skyrmions. The dipole fields
which point along the axis will add whereas the others will cancel in pairs.

Now consider a well separated charge 2 Skyrmion and a charge $-2$ antiSkyrmion 
with a common axis of symmetry and separation $s.$
If the Skyrmion is rotated by a phase $\alpha$ about the symmetry axis
then, because all the dipole fields lie along this axis, 
the leading order interaction energy will be negative and independent 
of $\alpha.$ To order $1/s^3$ the energy is given by
\be
E=2E_2-\frac{16C^2}{3\pi s^3}
\label{int2}
\ee
where $E_2$ is the energy of the charge 2 Skyrmion.

From the point of view of the interaction energy the charge 2 Skyrmion
and charge $-2$ antiSkyrmion pair is qualitatively similar to the 
monopole-antimonopole pair. The leading order contribution to the 
interaction is attractive and independent of the phase, so the pair
will not drift away to infinite separation. This configuration therefore
has more chance of forming a sphaleron than the charge 1 Skyrmion
and charge $-1$ antiSkyrmion pair, when only this aspect is considered.

As we discuss in detail in the appendix, 
there are good reasons to believe that for all $n>1$ 
the axially symmetric charge
$n$ Skyrmion has asymptotic fields that consist of only 
a single pion dipole, which is aligned with the symmetry axis,
and whose dipole strength increases with $n.$ Therefore the
previous discussion of the nature of the interaction energy of a
charge 2 Skyrmion and a charge $-2$ antiSkyrmion applies equally
well to a configuration consisting of an 
axially symmetric charge $n$ Skyrmion and a charge $-n$ antiSkyrmion
for all $n>1.$

Having seen that the dipole interaction between an axial
charge $n$ Skyrmion and a charge $-n$ antiSkyrmion is
favourable for the formation of a sphaleron let us
now consider the topology of such a configuration.

The fields of the axially symmetric  charge $n$ Skyrmion are invariant 
under a rotation by $2\pi/n$ around the symmetry axis. Note that although 
the Skyrmion is said to have an axial symmetry, generically 
a rotation by $\alpha$
around the axis of symmetry will change the fields, and 
axial symmetry refers to the fact
that this change is equivalent to an isospin rotation by $\alpha n.$ 

The closed loop that is relevant for an axial charge $n$ Skyrmion
and charge $-n$ antiSkyrmion pair is the creation of the pair from the vacuum, 
their separation, the rotation of the Skyrmion by $2\pi/n,$ and their
subsequent annihilation. The hope is then that a sphaleron would be
associated with the midpoint of this loop where the relative phase
is $\pi/n.$  
The closed loop which corresponds to the rotation by $2\pi/n$ around
the symmetry axis of an axially symmetric charge $n$ Skyrmion is 
non-contractible if and only if $n$ is odd \cite{Kr2}.
Thus the loop relevant for the sphaleron is only non-contractible if $n$
is odd. It therefore appears that the lowest value for which
the interaction energy and topology combine to produce a sphaleron
should be $n=3.$ 

Naively, there seems no reason
to suppose a sphaleron should exist with $n=2$ (or any other even value) 
since the topology appears to be lost, but as we shall see it turns out 
that this conclusion is too hasty.   
It may be that the energy barrier provided by the non-zero relative phase
is sufficient to yield a solution, or that the topology is more subtle.
For example, it may be the existence of non-trivial higher homotopy
groups that underlies the existence of these solutions. This is certainly
a possibility in the Skyrme model as there are non-contractible spheres
since
\be
\pi_2(\smon_0)
=\pi_2(\mbox{Maps}(S^3\mapsto S^3))=\pi_5(S^3)=\Z_2.
\label{cssky}\ee
Another possibility is that although for even $n$ the loop is
contractible in the full space of Skyrme fields it may not be 
contractible within the restricted space of axially symmetric
Skyrme fields. It would be interesting to clarify this, but we
have as yet been unable to do so. Should it be true then
symmetry considerations prevent an axially symmetric
field from breaking this symmetry during relaxation, so it should be
sufficient to yield a sphaleron.

In this section we have made some arguments for the possible existence
of sphalerons consisting of a charge $n$ Skyrmion
and charge $-n$ antiSkyrmion, where $n>1.$
 In the following sections we investigate this
using numerical methods and find that indeed such solutions exist.

\section{Axial Skyrmions}\news\label{sec-axial}
The static energy of the Skyrme model is given by
\be
E=\frac{1}{12\pi^2}\int \left\{-{1 \over 2}\mbox{Tr}(R_iR_i)-{1 \over 16}
\mbox{Tr}([R_i,R_j][R_i,R_j])\right\} \, d^3x\,,
\label{skyenergy}
\ee
where 
$R_i=(\partial_i U)U^\dagger$ is the $su(2)$-valued current
associated with the $SU(2)$-valued Skyrme field $U({\bf x}).$
With this normalization the Faddeev-Bogomolny bound is $E\ge |B|$
where the baryon number $B$ is
\be
B=-{1\over 24\pi^2}\int \epsilon_{ijk}
 {\rm Tr}\left(R_iR_jR_k\right)\, d^3x.
\label{baryon}
\ee 
 
To make contact with the nonlinear pion theory $U$ is written as
\be
U=\sigma +i\bpi\cdot\pauli
\ee
where $\pauli$ denotes the triplet of Pauli matrices,
$\bpi=(\pi_1,\pi_2,\pi_3)$ is the triplet of pion fields and $\sigma$
is determined by the constraint $\sigma^2+\bpi\cdot\bpi=1.$

In this paper we are only concerned with axially symmetric fields
so we introduce the ansatz
\be
\sigma=\psi_3,\quad
\pi_1=\psi_1\cos n\theta,\quad
\pi_2=\psi_1\sin n\theta,\quad
\pi_3=\psi_2
\label{ansatz}
\ee
where $\bpsi(\rho,z)=(\psi_1,\psi_2,\psi_3)$ is a three-component
unit vector which is a function only of $\rho$ and $z,$ where 
$\rho$ and $\theta$ are polar coordinates in the $(x_1,x_2)$ plane
and $z=x_3.$ In the above ansatz $n$ is an integer that counts the planar
winding of the fields.   

Substituting the ansatz (\ref{ansatz}) into the Skyrme energy 
(\ref{skyenergy}) gives
\be
E=\frac{1}{6\pi}\int
\{ (\partial_\rho\bpsi\cdot\partial_\rho \bpsi
+ \partial_z\bpsi\cdot\partial_z \bpsi)
(1+\frac{n^2}{\rho^2}\psi_1^2)
+|\partial_\rho\bpsi \times \partial_z\bpsi|^2
+\frac{n^2}{\rho^2}\psi_1^2
\}\ \rho\,d\rho\,dz
\label{axialenergy}
\ee  
which is a kind of Baby Skyrme model on the half-plane.
The baryon number is given by
\be
B=\frac{n}{\pi}\int \{
\psi_1 \bpsi\cdot \partial_\rho\bpsi \times \partial_z\bpsi\}
\ d\rho\,dz.
\ee

The finite energy boundary conditions are that
$\bpsi\rightarrow (0,0,1)$ as $\rho^2+z^2\rightarrow \infty,$
and on the symmetry axis $\rho=0$ we require $\psi_1=0$ and
$\partial_\rho\psi_2=\partial_\rho\psi_3=0.$

A configuration with the correct topology and boundary conditions 
of an axially symmetric Skyrme field with $B=n$ is given by
\be
\psi_1=\frac{\rho}{r}\sin f,\quad
\psi_2=\frac{z}{r}\sin f,\quad 
\psi_3=\cos f
\label{ic1}
\ee
where $r=\sqrt{\rho^2+z^2}$ and $f(r)$ is a monotonically
decreasing profile function with $f(0)=\pi$ and $f(\infty)=0.$

In order to create initial conditions for a charge $n$ Skyrmion and a 
charge $-n$ antiSkyrmion pair (which of course has total charge $B=0$) 
with separation $s$ 
we perform the following construction. Let $\bpsi^{(1)}$ be a 
configuration of the form (\ref{ic1}) with 
profile function $f(r)=\pi(1-2r/s)$ for
$r\le s/2$ and zero otherwise. Now make the replacement $z\mapsto z-s/2$ 
so that the charge $n$ Skyrmion is located at $z=s/2$ on the
symmetry axis. Let $\bpsi^{(2)}$ be a similar configuration but this
time shifted by  $z\mapsto z+s/2$  so that it is located at $z=-s/2.$
To turn this second Skyrmion into an antiSkyrmion we make the
reflection
\be
(\psi^{(2)}_1,\psi^{(2)}_2,\psi^{(2)}_3)\mapsto 
(\psi^{(2)}_1,-\psi^{(2)}_2,\psi^{(2)}_3)
\label{refl1}
\ee
which we refer to as an antiSkyrmion with zero relative phase, compared
to the original Skyrmion. Note that this reflection changes the sign
of only one of the pion fields and so our definition of the relative phase
differs by an addition of $\pi$ in comparison with 
that of refs.\cite{BGS,ILP,GP}, where the antiSkyrmion is obtained
by changing the sign of all three of the pion fields.
 However, the definition used in this paper makes for a more natural
comparison with the similar situation for monopoles. 

The fields of a charge $n$ Skyrmion are invariant under a rotation
around the symmetry axis by $2\pi/n,$ so the midpoint of the 
loop we are interested in corresponds to a rotation by $\pi/n.$
From (\ref{ansatz}) we see that the rotation $\theta\mapsto\theta+\pi/n$
changes the sign of the pion fields $\pi_1$ and $\pi_2$ and so 
is equivalent to the change $\psi_1\mapsto-\psi_1.$ Thus to 
obtain a charge $n$ antiSkyrmion which is out of phase with the
charge $n$ Skyrmion 
(in the sense of being at the midpoint of the loop) we replace
the transformation (\ref{refl1}) with 
\be
(\psi^{(2)}_1,\psi^{(2)}_2,\psi^{(2)}_3)\mapsto 
(-\psi^{(2)}_1,-\psi^{(2)}_2,\psi^{(2)}_3).
\label{refl2}
\ee    

Finally, the fields are defined to be
\be
\bpsi=\cases{
\bpsi^{(1)}\quad \mbox{ if}\quad \sqrt{(z-s/2)^2+\rho^2}\le s/2 \cr
\bpsi^{(2)}\quad \mbox{ if}\quad \sqrt{(z+s/2)^2+\rho^2}\le s/2 \cr
(0,0,1)\quad\quad  \mbox{otherwise.}
}
\label{patch}
\ee
The profile function is zero at a radius $s/2$ from the centre of the
Skyrmion (or antiSkyrmion), so the Skyrmion and antiSkyrmion
 are patched together
in a continuous way, with the fields set to the vacuum outside each
of the cores. This method of creating initial conditions is
preferable to using a product ansatz with a minimal energy profile 
function, since the product ansatz does not respect the symmetries
of the configuration in the same way that the patching ansatz (\ref{patch})
does.

In the following section we discuss the results of a numerical relaxation
of axially symmetric Skyrmion-antiSkyrmion pairs using the initial
conditions described above. In addition to the initial conditions using 
the simple ansatz (\ref{ic1}), more sophisticated  initial conditions
were also used based on the rational map ansatz \cite{HMS} with an
axially symmetric charge $n$ map. Either ansatz leads to the same
final solutions, though the rational map ansatz relaxes slightly faster
since it provides a better approximation to the axial charge $n$ Skyrmion.

\section{Numerical Results}\news\label{sec-num}
To find stationary points of the energy (\ref{axialenergy}) we solve
the associated gradient flow equation. This is an evolution equation
which is first order in a fictitious time and where the velocity of the
field is given by minus the variation of the energy, taking into account
the unit vector constraint. We do not present the details
of this equation here since it is rather cumbersome and not particularly
enlightening. The end point of the gradient flow evolution is then
the required stationary point of the energy.
The gradient flow equation is solved numerically using a finite difference
scheme which is second order accurate in the spatial derivatives 
and first order in the time derivatives. 
The grid in the $(\rho,z)$-plane contains
$200\times 400$ points with a lattice spacing of $0.05$ so that the range
covered is $(\rho,z)\in[0,10]\times[-10,10].$
\begin{figure}[ht]
\begin{center}
\leavevmode
\vskip -3cm
\epsfxsize=18cm\epsffile{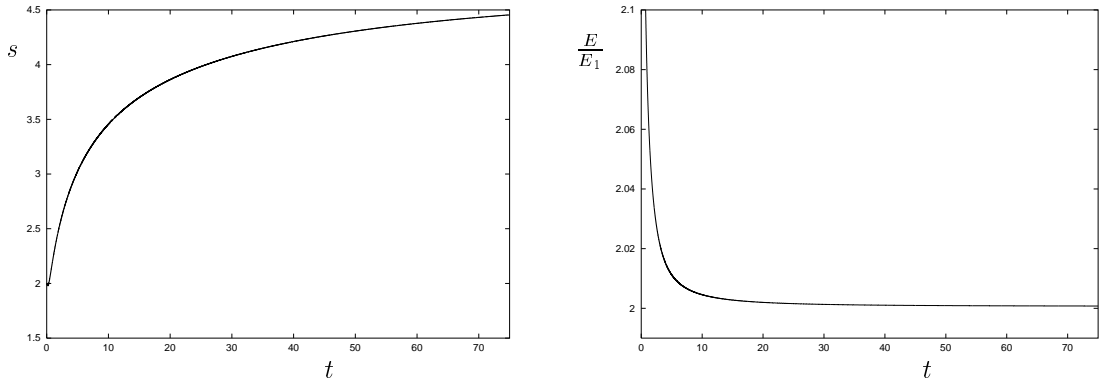}
\vskip -18cm
\caption{The separation $s$ and energy $E$ 
(in units of the energy of a single Skyrmion) as a function of time,
for a charge 1 Skyrmion and charge $-1$ antiSkyrmion pair with a
relative phase of $\pi.$ }
\label{fig-ch1}
\vskip -0.9cm
\end{center}
\end{figure}
\begin{figure}[hb]
\begin{center}
\leavevmode
\vskip -3cm
\epsfxsize=18cm\epsffile{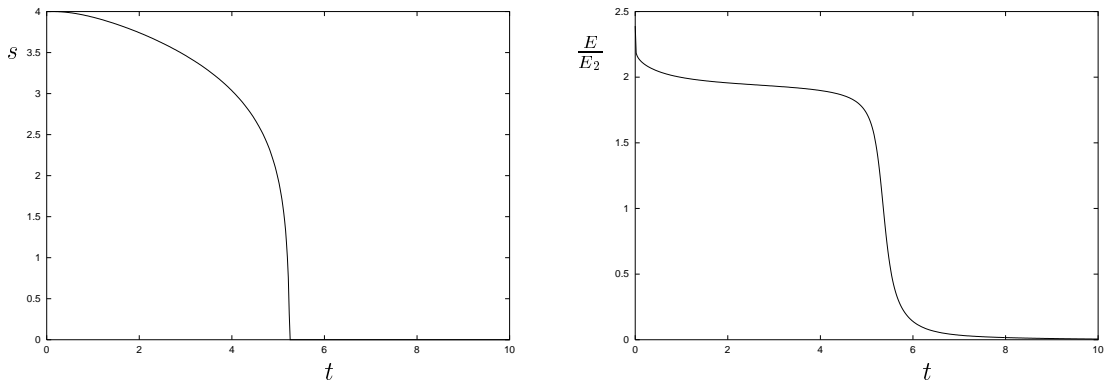}
\vskip -18cm
\caption{The separation $s$ and energy $E$ (in units of $E_2$)
 as a function of time,
for a charge 2 Skyrmion and charge $-2$ antiSkyrmion pair with zero
relative phase. }
\vskip -0.5cm
\label{fig-ch2a}
\end{center}
\end{figure}

As a test on the numerical code we first perform two simulations which
are not expected to lead to a sphaleron solution. The first simulation
consists of a charge 1 Skyrmion and charge $-1$ antiSkyrmion pair with a
relative phase of $\pi.$ The initial conditions are created using the 
ansatz described in the previous section with an initial separation $s=2.$
In Fig.~\ref{fig-ch1} we plot the separation $s,$ 
and the energy $E$ divided by the
energy of a single Skyrmion ($E_1=1.232$),
as a function of time during the gradient flow evolution.
The separation is computed from the positions of the Skyrmion and antiSkyrmion,
which are  defined to be the
points in space where the $\sigma$ field is equal to $-1.$
The separation increases with time and the 
energy tends towards twice the energy of
a single Skyrmion, confirming the repulsive force between this pair.

For the second simulation we turn to the charge 2 Skyrmion and charge
$-2$ antiSkyrmion pair. Fig.~\ref{fig-ch2a} displays the results of an initial
configuration with separation $s=4$ and zero relative phase. The separation
rapidly decreases to zero, as does the energy, demonstrating that the 
pair annihilate. The energy is plotted in units of $E_2,$ the energy of
the axially symmetric charge 2 Skyrmion. In this paper $E_n$ denotes the
energy of the axially symmetric charge $n$ Skyrmion, and energies will
be plotted in these units for ease of comparison. The values used for
$E_n$ can be found in the third column of Table~\ref{tab-energies}
 and were computed using
the same code with the same lattice values, so that an accurate computation
of the relative energies should result.
\begin{figure}[ht]
\begin{center}
\leavevmode
\vskip -2.5cm
\epsfxsize=18cm\epsffile{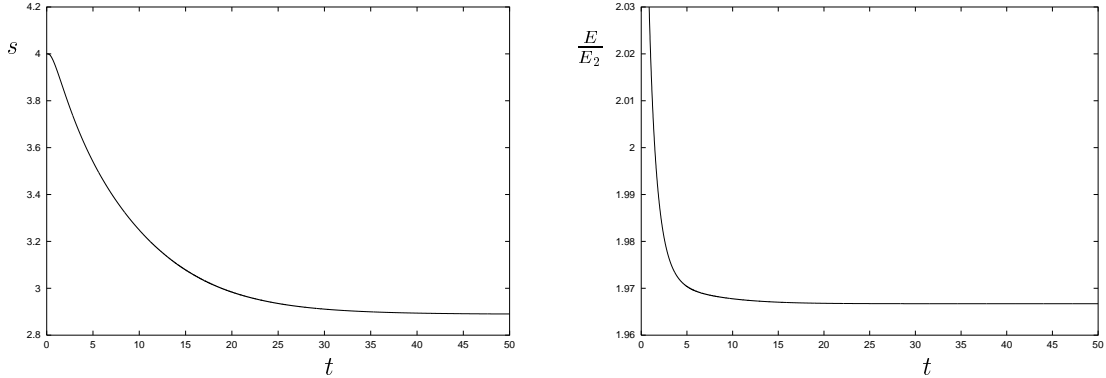}
\vskip -18cm
\caption{The separation $s$ and energy $E$ (in units of $E_2$)
 as a function of time,
for a charge 2 Skyrmion and charge $-2$ antiSkyrmion pair with 
relative phase $\pi/2.$ }
\label{fig-ch2s}
\vskip -0.5cm
\end{center}
\end{figure}

\begin{figure}[hb]
\begin{center}
\leavevmode
\vskip -0.5cm
\epsfxsize=10cm\epsffile{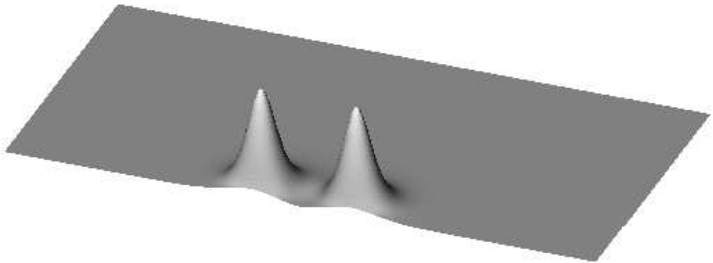}
\vskip -0cm
\caption{The sphaleron energy density in the $(z,\rho)$-plane.}
\label{fig-2d}
\end{center}
\end{figure}

Fig.~\ref{fig-ch2s} displays the results for a 
charge 2 Skyrmion and charge $-2$ antiSkyrmion pair
again with an initial separation $s=4$ but this time with a relative
phase of $\pi/2.$ As expected from the dipole analysis, 
the separation initially decreases but it then
tends to an asymptotic value $s=2.89,$ at which the energy is $1.967$
times the energy of the axially symmetric charge 2 Skyrmion ie.
the energy is slightly less than the energy of the Skyrmion-antiSkyrmion
constituents. This is the sphaleron solution. The sphaleron
energy density in the $(z,\rho)$-plane is displayed in Fig.~\ref{fig-2d}.
\begin{figure}[ht]
\begin{center}
\leavevmode
\vskip -2cm
\epsfxsize=14cm\epsffile{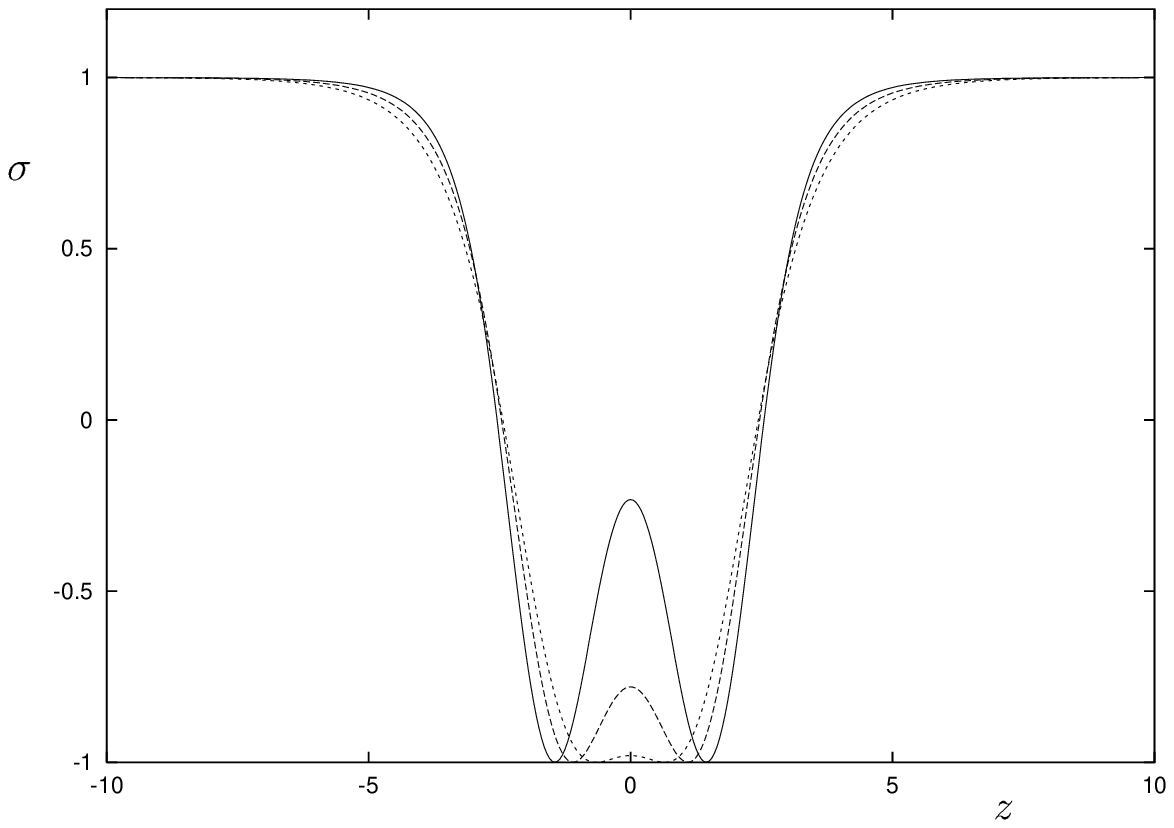}
\vskip -13cm
\caption{The $\sigma$ field along the $z$-axis for the sphaleron solution
formed from the charge $n$ Skyrmion and charge $-n$ antiSkyrmion, where
$n=2$ (solid curve), $n=3$ (dashed curve), $n=4$ (dotted curve).}
\label{fig-sigma234}
\end{center}
\end{figure}
The charge 2 Skyrmion and charge $-2$ antiSkyrmion are clearly visible 
as distinct structures and this explains why the energy is only very 
slightly less than twice the energy of the charge 2 Skyrmion.
The $\sigma$ field along the $z$-axis is plotted as the solid curve 
in Fig.~\ref{fig-sigma234}, where the soliton positions ($\sigma=-1$) 
can be seen on the $z$-axis at $z=\pm 1.445.$ 
Note that between the pair the field is far from the vacuum value
$\sigma=1.$

\begin{table}
\centering
\begin{tabular}{|c|r|c|c|c|}
\hline
$n$ & $E$\ \ \ & $E_n$ & $E/E_n$&$(z,\rho)$\\\hline
2 & 4.645 & 2.362 & 1.967 & $(\pm 1.445,0)$\\
3 & 6.947 & 3.581 & 1.940 & $(\pm 1.096,0)$\\
4 & 9.323 & 4.863 & 1.917 & $(\pm 0.653,0)$\\
5 & 11.656& 6.141 & 1.898 & $(0,1.303)$\\
6 & 14.054& 7.481 & 1.879 & $(0,2.082)$\\
\hline
\end{tabular}
\caption{For $2\le n\le 6$ the table shows the energy $E$ of the 
sphaleron consisting
of a charge $n$ Skyrmion and a charge $-n$ antiSkyrmion pair,
the energy $E_n$ of the axially symmetric charge $n$ Skyrmion, the 
ratio $E/E_n,$ and the position of the Skyrmion and antiSkyrmion in the
$(z,\rho)$-plane.}
\label{tab-energies}
\end{table}

Higher energy sphaleron solutions are formed from a charge $n$
Skyrmion and charge $-n$ antiSkyrmion pair with a relative phase
of $\pi/n,$ where $n>2.$ For $n=3$ and $n=4$ the sphaleron solution
is qualitatively similar to the case of $n=2.$ Table~\ref{tab-energies}
 lists the energies of these sphalerons, a comparison with the energies
of axial charge $n$ Skyrmions, and the 
positions of the Skyrmion-antiSkyrmion pair
in the $(z,\rho)$-plane. In Fig.~\ref{fig-sigma234} the $\sigma$ field 
along the $z$-axis is plotted as the dashed curve for $n=3$ and the dotted
curve for $n=4.$ Table~\ref{tab-energies} and Fig.~\ref{fig-sigma234}
demonstrate that the charge $n$ Skyrmion and charge $-n$ antiSkyrmion
 pair sit closer together and are more tightly bound as $n$ increases.
This is consistent with the fact that the dipole strength of the
axial charge $n$ Skyrmion increases with $n.$
Recall that, as discussed earlier, in the case $n=3$ the sphaleron
is associated with a non-contractible loop, since the relevant loop
is non-contractible when $n$ is odd. Thus it might be possible
to rigorously prove the existence of this $n=3$ sphaleron solution using
Taubes' approach.

\begin{figure}[ht]
\begin{center}
\leavevmode
\vskip -2cm
\epsfxsize=18cm\epsffile{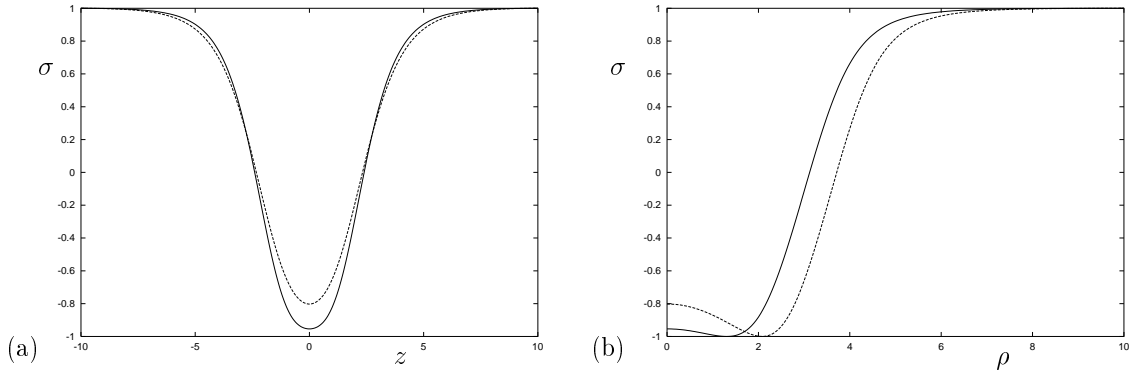}
\vskip -18cm
\caption{The $\sigma$ field along (a) the $z$-axis and
(b) the $\rho$-axis for the sphaleron solution
formed from the charge $n$ Skyrmion and charge $-n$ antiSkyrmion, where
$n=5$ (solid curves) and $n=6$ (dashed curves).}
\label{fig-sigma56}
\vskip -0.5cm
\end{center}
\end{figure}

For $n>4$ the sphaleron solution has an interesting qualitative difference
with the solutions described so far. The energies presented in 
Table~\ref{tab-energies} for $n=5$ and $n=6$ follow a similar trend
to the $n=2,3,4$ solutions. However, an examination of the $\sigma$ field
reveals that the positions of the Skyrmion and antiSkyrmion are no longer
two isolated points on the $z$-axis, but merge to form a circle. 
Fig.~\ref{fig-sigma56} plots the $\sigma$ field along (a) the $z$-axis
 and (b) the
$\rho$-axis, for $n=5$ (solid curves) and $n=6$ (dashed curves).
It can be seen that the $\sigma$ field never attains the value $-1$
along the $z$-axis, and in fact the minimal value along this axis
increases with $n.$ The $\sigma$ field is $-1$ on a circle in the $z=0$
plane of radius $\rho=1.303$ for $n=5$ and radius $\rho=2.082$ for $n=6.$
The Skyrmion and antiSkyrmion can no longer be identified, having
merged so that they are both located on a whole circle.
As we now describe, these solutions have some features in
common with Hopf solitons of the Skyrme-Faddeev model.

The field of the Skyrme-Faddeev model \cite{FN} is a three-component
unit vector, and it has topological soliton solutions classified by
the integer-valued Hopf invariant associated with the homotopy group
$\pi_3(S^2)=\Z.$ Solutions of the Skyrme-Faddeev model automatically
yield solutions of the Skyrme model by an embedding which sets one
of the pion fields to zero (say $\pi_3$) and maps the unit
vector onto the $\sigma$ field and the two remaining pion fields.
Such embedded solutions have zero baryon number and are expected to
be unstable, so they are sphaleron-like, but there is no obvious
interpretation in terms of Skyrmion-antiSkyrmion pairs.  

The lowest energy
Hopf soliton with Hopf charge one has a Skyrme energy of about $4.4$
(in the units we are using) so it has less energy than any of the
sphalerons presented in this paper. Axially symmetric Hopf solitons
exist for all Hopf charges $Q,$ although for $Q>2$ they are not the
minimal energy solitons and are unstable even within the
Skyrme-Faddeev model \cite{BS5}. Such axially symmetric solutions fit
into the class described by the ansatz (\ref{ansatz}) with $\psi_2=0$
and $Q=n.$ The $\sigma$ field is $-1$ on a circle and the pion fields
$\pi_1,\pi_2$ rotate $n$ times around this circle, so these solutions
are qualitatively similar to the sphaleron solutions formed from a
charge $n$ Skyrmion and charge $-n$ antiSkyrmion provided $n>4.$
The energies \cite{BS5} of the embedded axial Hopf solitons 
with $n=5$ and $n=6$
are $E=16.7$ and $E=19.9,$  so a comparison with 
Table~\ref{tab-energies} shows that it is not energetically favourable
to set $\psi_2=0$ for these values of $n$. 

We have seen that as $n$ increases the sphaleron solution has a $\sigma$ field
along the $z$-axis whose deviation from the vacuum value diminishes.
Embedded Hopf solitons have $\sigma=1$ along the entire $z$-axis, so it
may be that the sphaleron and Hopf soliton solutions tend towards the
same field configurations for large $n.$ It might be interesting to 
investigate this possibility in more detail in the future.   

\section{Conclusion}
In this paper we have discussed how considerations of asymptotic
pion dipole fields and topology combine to suggest the existence
of sphaleron solutions in the Skyrme model. These sphalerons
consist of an axial charge $n$ Skyrmion and an axial charge $-n$ antiSkyrmion 
(with $n>1$) balanced in unstable equilibrium. We have then used
numerical methods to compute these solutions and have described
their properties in some detail. The solutions appear to exists
for all $n>1,$ but the topological aspects are better understood for odd $n$,
and some open problems remain in clarifying the more subtle role of
the topology for even $n,$ though we have made some suggestions for
further investigation.

For $n>4$ a new interesting phenomenon occurs, with the position of
the Skyrmion and antiSkyrmion merging to form a circle, rather than
two isolated points. A similar phenomenon occurs in the Yang-Mills-Higgs
model, where charge $n$ monopole and charge $-n$ antimonopole solutions
also exist with $n\ge 1$ \cite{PT,KKS}, and for $n>2$ the Higgs field
is zero on a circle. This is another example of the similarity that 
is often found between monopoles and Skyrmions. There are also
monopole-antimonopole chains \cite{KKS2} where monopoles and
antimonopoles alternate along a line, and we expect similar solutions
to exist in the Skyrme model, provided the constituents are not single
Skyrmions or single antiSkyrmions.

Although the physical implications of saddle point solutions in
quantum field theory are not easily deduced, the existence of 
 sphaleron solutions in the Skyrme model may have 
ramifications which can be investigated experimentally. 
The simplest sphaleron solution we have found is relevant to 
deuterium-antideuterium annihilation, and although experiments can
not yet investigate this process, the recent experimental successes
in studying antihydrogen \cite{Amo} suggest it may be a possibility
in the near future.

Finally, the results presented in this paper may have technological
significance in future years, since apparently the engines of 
{\em Star Trek} starships are powered by deuterium-antideuterium reactors
\cite{SO}.

\section*{Acknowledgements}
Many thanks to Nick Manton and Tigran Tchrakian for useful discussions.
SK acknowledges the EPSRC for Research Fellowship GR/S29478/01.\\

\section*{Appendix}
In this appendix we discuss the asymptotic pion dipole fields of the
axially symmetric charge $n$ Skyrmion, where $n>1.$

From numerical calculations it is difficult to determine the 
asymptotic fields of a Skyrmion which is not spherically symmetric,
since a region is required which is both far from the Skyrmion core
and far from the boundary of the grid, and this is difficult to
achieve. Therefore, we present two approximate calculations
which both suggest that
for all $n>1$ 
the axially symmetric charge
$n$ Skyrmion has asymptotic fields that consist of only 
a single pion dipole, which is aligned with the symmetry axis,
and whose dipole strength increases with $n.$ 

Manton \cite{Ma13} has pointed out that it is often possible to predict the
asymptotic multipole fields of a given Skyrmion by simply adding together
the individual dipole fields of its constituents, if it is known how
to arrange the individual Skyrmions so that the required Skyrmion
will result. 

As we have discussed earlier, the fields of the axially symmetric charge
$n$ Skyrmion are strictly invariant under a rotation around the symmetry
axis through an angle $2\pi/n.$ It is possible to arrange $n$ well
separated single Skyrmions so that this cyclic subgroup is realized, and
it is likely that this is the appropriate alignment to yield the axially
symmetric charge $n$ Skyrmion. Similar cyclic arrangements of 
monopoles indeed lead to scattering processes which pass through the 
axially symmetric monopole \cite{HMM,Su5}.

Generically, a Skyrmion has a dipole as its leading multipole
\be
\pi_j=\frac{C_{ji}{\bf x}_i}{r^3}
\ee
where $4\pi C_{ji}$ is the dipole moment matrix. For a single 
Skyrmion we can choose the orientation so that
\be 
C_{ji}=\pmatrix{c&0&0\cr0&c&0\cr0&0&c}.
\ee

Arrange $n$ Skyrmions with positions
\be {\bf X}^{(k)}=(L\cos \frac{2\pi k}{n}, L\sin \frac{2\pi k}{n}, 0),
\quad\quad k=1,\ldots,n
\ee
where $L$ is a separation scale. If the orientations of the pion
fields are chosen to be
\be C^{(k)}=c\pmatrix{\cos({2\pi k}/{n})& -\sin({2\pi k}/{n})& 0\cr
\sin({2\pi k}/{n})&\cos({2\pi k}/{n})&0\cr
0 & 0&1\cr}
\ee
then this arrangement has cyclic symmetry, since under a spatial
rotation by $2\pi/n$ around the $x_3$-axis, Skyrmion $k$
is rotated into Skyrmion $k+1$ and $C^{(k)}$ maps to
$C^{(k+1)}.$

The sum of the dipole moment matrices gives
\be 
\sum_{k=1}^n C^{(k)}=\pmatrix{0&0&0\cr 0&0&0\cr 0&0&cn}
\ee
and so suggests that the axially symmetric Skyrmion has
asymptotic fields that consist of only 
a single pion dipole, which is aligned with the symmetry axis,
and whose dipole strength is proportional to $n.$
This argument is rather simple, and so the finer details should not
be trusted too much, but it seems likely that the qualitative picture is
correct, and for example the dipole field is likely to grow with
$n,$ though a simple linear growth is probably too simplistic. 

An approximate method for obtaining charge $n$ Skyrmions is through the
holonomy of a four-dimensional charge $n$ Yang-Mills instanton \cite{AM}.
For the axially symmetric Skyrmion the relevant instanton is of the
JNR type \cite{JNR} and is determined by a solution of the Laplace
equation of the form
\be
\zeta=\sum_{k=0}^{n}\frac{\lambda_k}{|x-a_k|^2}
\ee
where $x$ is the coordinate in four-dimensional Euclidean space, 
$\lambda_k=\frac{1}{n+1}$ are equal weights, and 
$a_{k\mu}=(L\cos \frac{2\pi k}{n+1}, L\sin \frac{2\pi k}{n+1}, 0,0)$
are pole positions with $L$ a length scale.

Defining the quantity 
\be Q_{\mu\nu}=\sum_{k=0}^n\lambda_k a_{k\mu}a_{k\nu} \ee
then the asymptotic dipole fields of the instanton generated Skyrmion
are given by \cite{LM}
\be
\pi_k=-\frac{\pi}{2}(Q_{11}+Q_{22}+Q_{33}-Q_{44})\frac{x_k}{r^3}
+(Q_{kj}-\varepsilon_{ijk}Q_{i4})\frac{\pi x_j}{r^3}.
\ee
Substituting the above values yields
\be
\pi_k=-\frac{L^2 \delta_{k3}\pi x_3}{2r^3}
\ee
which again suggests only a single pion dipole field aligned with
the symmetry axis. Note that the sign is irrelevant here since there
is only one non-zero component so its sign
can be changed by an isospin transformation. 
The scale $L$ needs to be fixed in the instanton
approximation by a minimization over $L$ of the energy of the resulting 
approximate Skyrmion. Since the size of an axially
symmetric Skyrmion grows with $n$ it is expected that so does the 
minimizing scale $L,$ so again the prediction is that the dipole
strength should increase with $n,$ in broad agreement with the 
prediction of the first approach.

\end{document}